*Article*

# Identifying Heart Attack Risk in Vulnerable Population: A Machine Learning Approach

Subhagata Chattopadhyay [1], and Amit K Chattopadhyay [2,*]

1. Institute of Health Management Research, Electronic City, Bangalore 560105, Karnataka India; subhagata.vf@iihmrbangalore.edu.in
2. School of Business, National College of Ireland, Mayor Street Lower, Dublin 1, Ireland; amit.chattopadhyay@ncirl.ie
* Correspondence: amit.chattopadhyay@ncirl.ie; Tel.: +353 1800 221 721

**Abstract:** The COVID-19 pandemic has significantly increased the incidence of post-infection cardiovascular events, particularly myocardial infarction, in individuals over 40. While the underlying mechanisms remain elusive, this study employs a hybrid machine learning approach to analyze epidemiological data in assessing 13 key heart attack risk factors and their susceptibility. Based on a unique dataset that combines demographic, biochemical, ECG, and thallium stress-tests, this study aims to design, develop, and deploy a clinical decision support system. Assimilating outcomes from five clustering techniques applied to the 'Kaggle heart attack risk' dataset, the study categorizes distinct subpopulations against varying risk profiles and then divides the population into 'at-risk' (AR) and 'not-at-risk' (NAR) groups using clustering algorithms. GMM algorithm outperforms its competitors (clustering accuracy and Silhouette coefficient scores are 84.24% and 0.2623, respectively). Subsequent analysis, employing Pearson correlation and linear regression as descriptors, reveals strong association between the likelihood of experiencing a heart attack on the 13 risk factors studied and these are statistically significant ($p < 0.05$). Our findings provide valuable insights into the development of targeted risk stratification and preventive strategies for high-risk individuals based on heart attack risk scores. The aggravated risk for postmenopausal patients indicates compromised individual risk factors due to estrogen depletion that may be, further compromised by extraneous stress impacts, like anxiety and fear, aspects that have traditionally eluded data modeling predictions. The model can be repurposed to analyze the impact of COVID-19 in vulnerable population.

**Keywords:** Heart attack; Risk prediction; CDSS; Clustering; COVID-19; Linear regressions; Pearson's correlation.

## 1. Introduction

A growing body of evidence suggests a significant increase in post-COVID-19 cardiovascular events, particularly myocardial infarction, with disproportionate impact on younger populations. While cardiovascular diseases remain a leading cause of global mortality, claiming approximately 20.5 million lives annually [1], the post-COVID-19 era has witnessed a concerning rise in cardiac events among individuals under 40 [2].

This phenomenon may be attributed to a complex interplay of factors, including post-acute COVID-19 sequelae such as persistent infection, post-COVID-19 syndrome (long COVID), and long-haul COVID-19 [3]. Furthermore, potential contributions from vaccine-related side effects, including myocarditis, dysrhythmias, and thromboembolic events, cannot be excluded [4].

The pathophysiological mechanisms linking COVID-19 to cardiovascular complications are multifaceted and include: (i) direct viral damage to vascular endothelium, leading to impaired blood flow; (ii) persistent viral reservoirs in various tissues; and (iii) dysregulation of the immune system. These factors may collectively contribute to an increased risk of adverse cardiovascular outcomes across all age groups [3]. The virus has a known affinity towards the cardiovascular system with a biochemical pathway connecting the pulmonary system. This pathway connection is perceived as a key player in exacerbating heart attacks and is seen as one of the many manifestations of COVID-19 impact on cardiovascular diseases (CVD) [5]. A recent study shows acute coronary disease, myocardial infarction (heart attack), angina, and ischemic cardiomyopathy (due to vasculopathy) as predominant causes of heart attack irrespective of age groups [6].

This analyzes temporal data patterns to predict the risk of heart attacks, irrespective of COVID-19 infections. It is interesting to note that despite having vulnerabilities due to exposure from well-established causative factors, not all patients are equally susceptible to heart attacks. Profiling patients according to their CVD susceptibility is a challenging area in healthcare analytics which is the premise of this study. It is therefore important and pertinent to understand the electrophysiology of the heart. The *heart* is an autonomous organ, a metronome that beats seamlessly in its way and is influenced by several physiological factors, e.g., autonomic stress (high norepinephrine drive), oxidative stress (super-oxide-led destruction of vascular tissues), emotional stress (anxiety disorders) leading to sympathetic overdrive, and many others [7]. The heart relies on a delicate balance between sympathetic and parasympathetic activity, modulated by the vagus nerve [8] to maintain a state of physiological equilibrium. This dynamic is reflected in several Heart rate variability (HRV) metrics such as SDNN, pNN50, RMSSD, and others [9]. Electrophysiological abnormalities, including atrial fibrillation, ectopic beats, and missed beats, contribute to HRV patterns, with a threshold beyond which they signify pathological conditions [10]. Consequently, cardiac function is intricately linked to both electrophysiological and electrophysiological factors, influencing the HRV outcomes. This complex interplay contributes to the variability in individual responses to cardiac risk factors, making accurate prediction of myocardial infarction a significant clinical challenge. Given this background, this article presents a novel Machine Learning-(ML) based analytical framework for cardiovascular risk assessment to analyze the heart attack risk from longitudinal data. This framework has the potential to significantly improve patient outcomes by enabling early diagnosis and personalized risk stratification.

Unfortunately, the *literature* on heart-attack risk prediction is not particularly well-formed. Sporadic studies have reported (a) longitudinal risk predictions over a span of ten years where the prediction methodology is largely rule-based and hence rigid [11], (b) data mining methods for exploratory data analysis to identify the key predictors of heart attack risk [12], and (c) risk scoring to prioritize the high-risk sub-population within a population with cardiovascular diseases (CVD) [13]. Thus, a key *objective* of this article is to propose a hybrid Machine Learning approach to design, develop, and deploy a clinical decision support system (CDSS) to (i) identify the CVD-vulnerable population using a combination of unsupervised machine learning (clustering) models and (ii) supervised model, such as linear regressions for scoring the risk. The study also compares the performances of different clustering techniques on heart-attack risk datasets, drawn from the Kaggle public data source [14]. It is important to note that the Kaggle heart risk data was made available in the public domain in 2021. From the data and its background information, it is unclear whether the impact of COVID-19 on the cardiovascular system has been considered. This is a lateral target of this study, namely, to associate COVID-19 connection with Kaggle data based on statistical modeling.

The *aim* of this study is to identify significant predictors within the high-risk clusters (identifying the best subjective clustering technique as an inclusive challenge) using Pearson's correlation coefficient values as the key descriptor and leverage these values to form a linear regression model for scoring the risk level for each case. On the other hand, the objective of the study is to design, develop, and deploy a hybrid ML model using a combination of unsupervised and supervised learning algorithms to not only predict the risk but also its associated score. The predicted ML scores are expected to complement the human evaluation by the clinicians in identifying the high-risk population, that will both be non-invasive and preemptive. It would help clinicians and nurses to prioritize the cases.

## 2. Materials and Methods

The details of the Kaggle heart data $303 \times 13$)($303 \times 13$) are available from [14]. Table 1 shows first 10 rows as the sample dataset.

**TABLE 1**
**A SAMPLE DATASET WITH TARGET AND ATTRIBUTES (N=10)**

| No. | age | sex | cp | trtbps | chol | fbs | restecg | thalachh | exng | oldpeak | slp | caa | thall | output |
|---|---|---|---|---|---|---|---|---|---|---|---|---|---|---|
| 1. | 63 | 1 | 3 | 145 | 233 | 1 | 0 | 150 | 0 | 2.3 | 0 | 0 | 1 | 1 |
| 2. | 37 | 1 | 2 | 130 | 250 | 0 | 1 | 187 | 0 | 3.5 | 0 | 0 | 2 | 1 |
| 3. | 41 | 0 | 1 | 130 | 204 | 0 | 0 | 172 | 0 | 1.4 | 2 | 0 | 2 | 1 |
| 4. | 56 | 1 | 1 | 120 | 236 | 0 | 1 | 178 | 0 | 0.8 | 2 | 0 | 2 | 1 |
| 5. | 57 | 0 | 0 | 120 | 354 | 0 | 1 | 163 | 1 | 0.6 | 2 | 0 | 2 | 1 |
| 6. | 57 | 1 | 0 | 140 | 192 | 0 | 1 | 148 | 0 | 0.4 | 1 | 0 | 1 | 1 |
| 7. | 56 | 0 | 1 | 140 | 294 | 0 | 0 | 153 | 0 | 1.3 | 1 | 0 | 2 | 1 |
| 8. | 44 | 1 | 1 | 120 | 263 | 0 | 1 | 173 | 0 | 0 | 2 | 0 | 3 | 1 |
| 9. | 52 | 1 | 2 | 172 | 199 | 1 | 1 | 162 | 0 | 0.5 | 2 | 0 | 3 | 1 |
| 10. | 57 | 1 | 2 | 150 | 168 | 0 | 1 | 174 | 0 | 1.6 | 2 | 0 | 2 | 1 |

The dataset was uploaded around 2021 when the second wave of COVID-19 and vaccination drives were in full swing. Hence, the influence of COVID-19 with heart health risk has been assumed in this work despite no direct link between COVID-19 pandemic and/or COVID-19 vaccination with the heart health dataset has been found. It is also worth mentioning that the attributes of the dataset display a unique combination of demographic, biochemical, ECG, and thallium stress-test data which have been considered in this work. For running clustering algorithms, initially, the dependent predictor (called 'outcome') with the class labels ('0' as not at risk or NAR and '1' as at risk or AR) is removed from the dataset. However, later, the performances of the algorithms are tested matching the AR and NAR labels.

*Summary of Kaggle heart risk dataset:* The shape of the dataset is $303 \times 13$ ($303 \times 13$) meaning 303 cases each having 13 predictors or attributes and one target variable, i.e., the outcome label (AR or NAR). As mentioned, initially, to run the clustering algortihm, the outcome labels are removed from the dataset. The features within the dataset are (i) *Age* (true value), (ii) *Sex* (male and female), (iii) Chest pain or *cp* ('0' typical angina, '1' atypical angina, '2' non-anginal pain, '3' asymptomatic), (iv) Resting blood pressure in mmHg (*trtbps*), (v) Cholesterol in mg% (*chol*), (vi) Fasting blood sugar (FBS) in mg% >120 mg% is '1' and '0' otherwise, (vii) Resting ECG (*restecg*) presented as '0' is normal, '1' is normal ST-T wave, and '2' denotes left ventricular hypertrophy, (viii) Maximum heart rate achieved (*thalachh*) while on Thallium stress test, (ix) Previous peak (*oldpeak*), (x) Slope (*slp*), (xi) Number of major vessels involved (*caa*), (xii) Thallium stress test results (*thall*) distributed between 0 and 3 based on the severity level, (xiii) Exercise induced angina (*exng*) presented as '0' no and '1' yes.

The dataset's two clusters correspond to the AR group (labeled '1', with 165 instances) and the NAR group (labeled '0', with 138 instances), respectively.

Prior to clustering, the dataset underwent tests for internal consistency and normality. Cronbach's alpha [15] was calculated to assess the reliability of the data. An alpha value above 0.8 indicates high internal consistency. The Shapiro-Wilk test [16] was used to evaluate the normality of the data, with a p-value greater than 0.05 (CI 95%) suggesting a normal distribution. In this case, the alpha value was found to be 0.08, indicating poor internal consistency, while the Shapiro-Wilk test returned a p-value of 0, indicating that the data were not normally distributed. Such results are often encountered with clinical and biological datasets.

This section presents an analysis of five clustering techniques employed to identify AR populations, along with their performance in correctly classifying the data. The ML models have been implemented using 5-fold cross validation technique with all variations in heart attack risk scores duly noted.

*Clustering technique-1:* The k-means clustering (k-MC) algorithm [17] is an unsupervised machine learning technique that requires the predefined number of clusters, denoted as 'k.' The algorithm operates without the need for training and iteratively identifies cluster centroids until convergence. Convergence is achieved when the algorithm has processed the entire dataset, ensuring that no data point remains unassigned to a cluster and the loss function is the minimum. Once the centroids are determined, the algorithm computes the distances between the centroids and other data points, assigning each data point to the cluster with the nearest centroid to minimize the sum of distances. This process is iterated to form all clusters. In the present study, the number of clusters, *k*, was predefined as 2, representing the AR and NAR groups.

*Clustering technique-2:* The Gaussian Mixture Model (GMM) technique operates similarly to k-means clustering (k-mc); however, it offers three distinct advantages. First, GMM accounts for variance, which corresponds to the horizontal spread or width of the normal distribution curve, thereby providing a more nuanced representation of the data. Second, GMM is well-suited for non-circular datasets, where data points are not measured in circular units such as radians [18]. Third, GMM employs a soft clustering approach, unlike the hard-clustering approach of k-mc. In k-means, each data point is assigned exclusively to a single cluster, with no consideration for its likelihood or belonging to other plausible clusters. By contrast, GMM assigns probabilities to data points for inclusion in multiple clusters, enhancing its flexibility and robustness in cluster analysis [19].

*Clustering technique-3:* The Density-Based Spatial Clustering of Applications with Noise (DBSCAN) algorithm classifies data points based on density variations, effectively distinguishing regions of higher density from those of lower density [20]. This algorithm is built on the foundational concepts of clusters and noise, requiring a minimum number of data points within a specified radius around each point to form a cluster. Unlike partition-based clustering methods, which generate spherical-shaped clusters and are highly sensitive to noise and outliers, DBSCAN excels in handling datasets with irregular cluster shapes and noise - a common characteristic of real-world data. DBSCAN relies on two critical parameters: (a) ε (epsilon), which defines the neighborhood radius, and (b) **MinPts**, the minimum number of points required to form a dense region referring to a cluster. The selection of these parameters significantly impacts clustering performance. If ε is set too large, data points may collapse into a single cluster, while an excessively small ε results in most points being categorized as outliers. To optimize ε, a k-distance graph can be employed to identify an appropriate threshold. The value of **MinPts** depends on the data dimensionality; for two-dimensional datasets, a typical choice is **MinPts = 2 + 1 = 3** or higher. This parameter tuning is crucial for accurately capturing the underlying structure of the dataset.

*Clustering technique-4: The Balance iterative reducing and clustering using hierarchies (BIRCH)* segregates the data points into summaries which are then clustered instead of clustering the data points [21]. Hence, it works better on large datasets compared to others. Each summary possesses as much information of the summaries as necessary and that is why it is incorporated in this work. It works well with numerical data. Hence, categorical data values need to be transformed into numerical ones to run BIRCH algorithm.

*Clustering technique-5:* The Affinity Propagation (AP) clustering algorithm identifies clusters based on the "affinity" or similarity between data points, selecting specific points, termed 'exemplars' to represent each cluster. Exemplars are determined through an iterative process that maximizes the overall similarity within clusters. Notably, akin to the DBSCAN algorithm, Affinity Propagation does not require prior specification of the number of clusters, offering flexibility in analyzing datasets with unknown cluster structures [21].

It is important to note that the work proposes designing, developing, and deploying a CDSS where clusters are chosen for grouping the sample into 'AR' and 'NAR'; the maximally overlapping cluster is chosen as the best cluster. Other performance metrics, e.g., compactness and distances between a pair of clusters, have thus been disregarded.

In the following section, Figures 2a-e depict the clustering performance. The performance is measured on the percentage of samples, correctly clustered. The high-risk cluster information is then engineered to find the most common predictors of heart attack risk and finally, they are correlated to compute the correlation coefficients using Pearson's correlation technique [8] (refer to equation 1) as a mechanism of Exploratory data analytics (EDA),

$$r = \sum(x_i - x')(y_i - y') / \sqrt{\sum(x_i - x')^2 \sum(y_i - y')^2} \quad \ldots (1)$$

In Equation (1), the parameter '*r*' represents the correlation coefficient, quantifying the strength and direction of the linear relationship between two predictors. The coefficient *r* ranges from −1 to +1, where 'positive' values indicate direct correlation, 'negative' values represent an inverse correlation, and 'zero' denotes no correlation. The terms $x_i$ and $y_i$ correspond to the i$^{th}$ sample values (with *i* ranging from 1 to *N*), while *x'* and *y'* denote the mean values of the respective variables. Positive '*r*' values, particularly those on the AR spectrum, are interpreted as significant predictor coefficients. The corresponding p-values < 0.05 indicate statistically significant correlations and ensure linearity in the model. These high '*r*' values are then leveraged to form a linear regression equation [22] for heart attack risk scoring (Equation (2)) as cluster information is not necessarily easy to translate into numbers,

$$y = \sum x_i b_i + c \quad \ldots (2)$$

In this equation 'y' is the predicted output (the heart attack risk scores), '$b_i$' is the i$^{th}$ coefficient value (where 'i' varies between 1 to N), and 'c' is the constant which is set at '0' (straight line passing through the origin in a least square fit). It is important to note that 'b' values are replaced by high 'r' values in this work to compute the respective 'y'. Here '$x_i$' denotes the i$^{th}$ ~~case~~ attribute for which 'y' is predicted for any given case.

Therefore, a hybrid ML model (clustering plus linear regressions) for heart attack risk scoring with EDA (Pearson's correlation coefficients, i.e., the 'r' values) to understand the significant predictors of heart attack risk is thus proposed, developed, and deployed in this work. All coding was done with Python 3.11.1 on a Windows 64-bit system having 8 GB RAM and Intel(R) Core (TM) i5-6300U CPU @ 2.40GHz.

## 3. Results

Descriptive statistics and data distributions for these features can be found in Table 2. It shows the description of the whole dataset, its central tendency, and its distribution.

TABLE 2
DESCRIPTIVE STATISTICS OF ALL VARIABLES (Count 303, M: 207, F: 96)

| Parameter | age | cp | trtbps | chol | fbs | restecg | thalachh | exng | oldpeak | slp | caa | thall |
|---|---|---|---|---|---|---|---|---|---|---|---|---|
| Mean | 54.3 | 0.68 | 0.96 | 131.62 | 246.26 | 0.14 | 0.52 | 149.64 | 0.32 | 1.03 | 1.39 | 0.72 |
| Std | 9.08 | 0.46 | 1.03 | 17.53 | 51.83 | 0.35 | 0.52 | 22.90 | 0.46 | 1.16 | 0.61 | 1.02 |
| Min | 29 | 0 | 0 | 94 | 126 | 0 | 0 | 71 | 0 | 0 | 0 | 0 |
| 25% | 47.5 | 0 | 0 | 120 | 211 | 0 | 0 | 133.5 | 0 | 0 | 1 | 0 |
| 50% | 55 | 1 | 1 | 130 | 240 | 0 | 1 | 153 | 0 | 0.8 | 1 | 0 |
| 75% | 61 | 1 | 2 | 140 | 274.5 | 0 | 1 | 166 | 1 | 1.6 | 2 | 1 |
| Max | 77 | 1 | 3 | 200 | 564 | 1 | 2 | 202 | 1 | 6.2 | 2 | 4 |
| Skewness | -0.2 | -0.8 | 0.5 | 0.7 | 1.14 | 2.0 | 0.16 | -0.5 | 0.7 | 1.26 | -0.5 | 1.3 |
| Kurtosis | -0.5 | -1.4 | -1.2 | 0.9 | 4.5 | 2.0 | -1.36 | -0.06 | -1.46 | 1.58 | -0.63 | 0.84 |

Table 2 shows how each parameter is distributed and the central tendency of the sample. Here 'Std', 'Min', 'Max', '25%', '50%','75%' refer to standard deviation, minimum value, maximum value, 1st, 2nd, and 3rd quartile, respectively. Skewness and Kurtosis values respectively determine the skewed deviation of the distribution curves and the thickness of the tail of the curves, predictor-wise.

Below, the relevant performance ~~screenshots~~ of the clustering techniques are shown and discussed. The performance metrics are a) classification accuracy (equation 3) and b) Silhouette score (equation 4). The score is a measure of compactness of datapoints within a cluster and how well the clusters are separated from each other).

$$A = \frac{1}{N} \sum val(AR) + val(NAR) \quad \dots (3)$$

In this equation, 'N' is the sample size, 'val(AR)', and 'val(NAR)' refers to validated AR and NAR cases respectively which are accurately clustered.

$$S(i) = \frac{b(i) - a(i)}{\max(a(i), b(i))} \dots (4)$$

Here, a(i) denotes the average distance between an observation and the datapoints within the same cluster and b(i) refers to the average distance between an observation and other datapoints in the nearest cluster. In a good clustering algorithm, we expect the distances are appreciable to separate the datapoints as well as the clusters.

The *k-mc clustering technique* (Silhouette score = 0.1782) returns the following result: Out of 165 samples under the cluster AR, 55 have been correctly clustered, which is 33.33%; while 78 samples are correctly clustered under NAR, which is 56.52% with an *average accuracy of 44.92%*. Hence, 55.08% are misclassified.

The *GMM clustering technique* (Silhouette score = 0.2623) shows the following result (Figure 1a): Out of 165 samples under the cluster AR, 139 have been correctly clustered, which is 84.24% and misclassified samples (which is 15.76%) are removed from further study; while 123 samples are correctly clustered under NAR (N=138), which is 89.13% with an *average accuracy of 86.68%*. Hence, 13.53% are treated as outliers. Fig. 1a shows the cluster plots, where green dots (cluster 1) refer to the AR group. It is also evident that the clusters are quite distinctly separated.

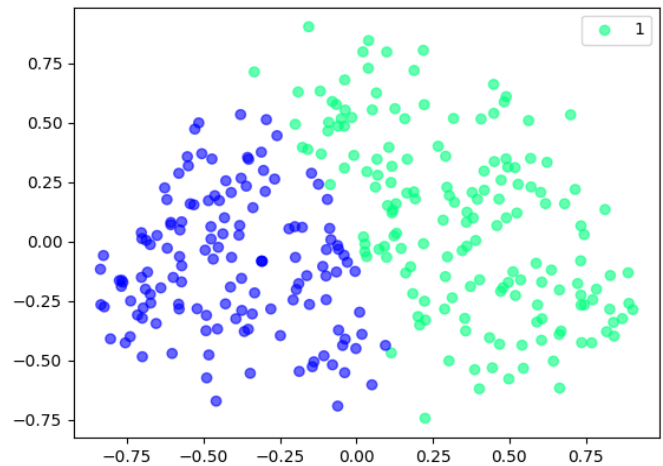

**Figure 1a.** GMM-based cluster plots.

The *DBSCAN clustering technique* (Silhouette score = NA) shows the following result after parametrizing the most appropriate ε value by computing k-distances (see Figure 1b), where the 'x' and 'y'-axis represent the datapoints and k-distances, respectively. The highlighted zone indicates 'ε' at 2.6, where the elbow is crooked at the first instance.

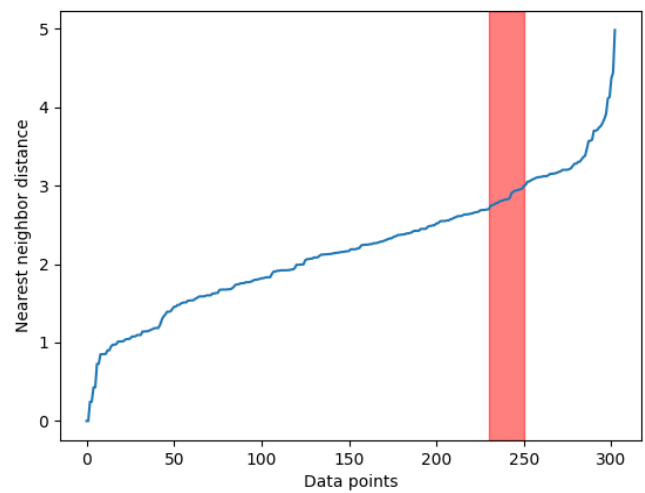

**Figure 1b.** Parametric study to find the best 'ε' value

The parameter 'MinPts' is chosen as twice the number of data dimensions, as recommended in prior studies [23]. For the Kaggle heart-attack risk dataset, which comprises 13 predictors, 'MinPts' is calculated as 13 × 2 = 26, corresponding to the 26 nearest neighbors. Subsequently, the distance to the (MinPts−1) nearest neighbors is computed for each data point within the dataset. These distances are then ranked in ascending order to generate a characteristic "elbow" plot [23]. The optimal value of 'ε' is shown along the y-axis value (Figure 1b). The evaluation is based on the mean of 5 nearest neighbor distances at the most prominent inflection point of the elbow (marked zone in Fig. 1b), which, in this case, is 2.6. With 'ε' and 'MinPts' values of 2.6 and 26, respectively, DBSCAN clustering algorithm is run to obtain the best clusters. In DBSCAN, all the data points are considered outliers (see Figure 1c).    Hence, it is excluded from cluster comparisons.

The *BIRCH clustering technique* (Silhouette score = 0.1509) shows the following result: Out of 165 samples under the cluster AR, 40 are correctly clustered, which is 24.24%; while, 80 samples are correctly clustered under NAR, which is 57.97% with an average accuracy of *41.1%*. Hence, 58.89% are misclassified.

The *Affinity propagation clustering algorithm* (Silhouette score = 0.1738) shows the following results: Out of 165 samples under the cluster AR, 104 are correctly clustered, which is 63.03%; while, 69 samples are correctly clustered under NAR, which is 50% with an average accuracy of *56.5%*. Hence, 43.5% are misclassified.

A summary of the clustering performance can be seen in Table 3.

**TABLE 3**
SUMMARY OF THE CLUSTERING PERFORMANCE (AVERAGE)

| No. | Technique | Correctly classified AR cases (%) | Silhouette score |
|---|---|---|---|
| 1 | k-mc | 33.33 | 0.1782 |
| 2 | *GMM* | ***84.24*** | ***0.2623*** |
| 3 | DBSCAN | NA | NA |
| 4 | BIRCH | 24.24 | 0.1509 |
| 5 | Affinity propagation | 63.03 | 0.1738 |

Key observations and advantages of clustering:

- The GMM technique demonstrated superior performance based on both the accuracy and Silhouette score coefficient measures, identifying 84.24% of high-risk cases with a score of 0.2623. As a result, GMM-based clusters were selected for visualization and further study.
- In clinical decision-support systems, the accurate identification of the AR cases is critical, as failure to detect such cases could lead to severe consequences, including loss of life. Among the five clustering techniques evaluated, *GMM emerged as the most clinically reliable approach* though its Silhouette score indicates some chances of misclassifications and thus overlapping between AR and NAR groups.
- The cluster information derived from GMM provides significant predictors with high correlation coefficients (r' values). These predictors were subsequently utilized to develop a linear regression model for calculating heart risk scores for individual cases.
- Computed heart risk scores prioritize vulnerable patients, facilitating effective management and timely intervention strategies.

Results of *Pearson's correlation* on the GMM-based high-risk cluster (see Table 4 and Fig. 2):

**TABLE 4**
SIGNIFICANT COEFFICIENTS (THE HIGHEST 'r' VALUES)

| No. | Predictor 1 | Predictor 2 | r | p-value |
|---|---|---|---|---|
| 1 | Age | trtbps | 0.28 | < 0.05 |
| 2 | Sex | Thall | 0.21 | |
| 3 | cp | thalachh | 0.30 | |
| 4 | trtbps | oldpeak | 0.19 | |
| 5 | chol | Age | 0.21 | |
| 6 | fbs | caa | 0.14 | |
| 8 | *thalachh* | *slp* | *0.39* | |
| 9 | exng | oldpeak | 0.29 | |
| 10 | oldpeak | caa | 0.22 | |
| 11 | caa | Age | 0.27 | |

| 12 | thall | oldpeak | 0.21 |

From Table 4, it is evident that the predictors 'slp' and 'thalachh' have the maximum correlation value of 0.38, hence any of 'slp' or 'thalachh' can be chosen for the model. Here, 'thalachh' has been chosen, and 'slp' is ignored as it has no other meaningful correlations with any of the remaining predictors, unlike 'thalachh', which also shows a high positive correlation with 'cp' (0.29). All the above 'r' values are now leveraged to form the linear regression equation (see equation 2).

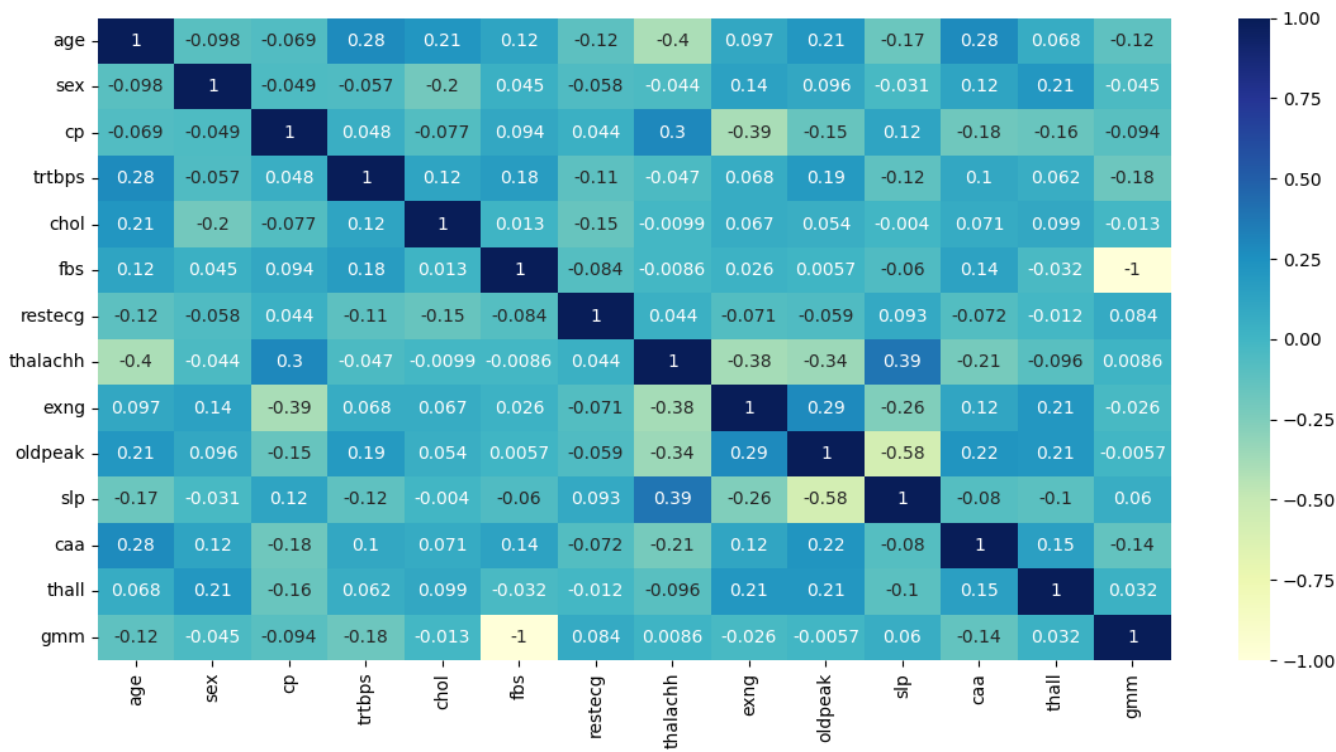

**Figure 5.** Pearson's correlation coefficient heatmap

Since the 'r' values are not sufficiently high for independent validation, these are tested for statistical significance. The p-values fall much below 0.05 and hence the linearity among the variables is statistically significant.

A detailed hand calculation is provided below for Case 1, while the heart attack risk scores for the top three high-risk cases are summarized in Table 5.

i) Predictor values, including Age ($x_1$), Sex ($x_2$), Chest Pain Type ($x_3$), …, and Thalassemia ($x_{12}$) are extracted from the Kaggle Heart Attack Risk dataset.
ii) Corresponding regression coefficients (r' values) are obtained from Table 4.
iii) Each predictor value is multiplied by its respective 'r' value (refer to Table 4).
iv) The resulting products are added to calculate the predictors.
v) A constant term, assumed to be zero in this analysis, is added to the computed sum, yielding the final heart attack risk score, in accordance with Equation (2).
vi) For Case 1, the calculated risk score (y) is 90.89.

This procedure is systematically applied to all cases in the dataset to compute their respective heart attack risk scores.

In the AR cluster, the *heart risk scores* are distributed as in Table 5. In this context, it is important to note that the remaining 13.53% of cases that have been misclassified by the GMM technique are also scored for heart attack risks. However, their values are less than the top three cases and therefore have not been presented in Table 5.

**TABLE 5**
THE DISTRIBUTION OF HEART RISK SCORES ACROSS HIGH-RISK CLUSTER OF GMM TECHNIQUE

| Parameter | Value |
|---|---|
| Count | 139 |
| Min | 90.89 |
| Max | 168.12 |
| Mean | 121.17 |
| Median | 122.02 |
| Standard deviation | 10.97 |
| Skewness | 0.45 |
| Kurtosis | 2.27 |

From the values of skewness and kurtosis, it is evident that the high-risk cases are symmetrically distributed as these values lie between [-2,2] and [-7,7] [24].

Individual heart attack risk score is computed using Equation (2) and showcased in Table 6.

**TABLE 6**
TOP THREE CASES WITH PREDICTED AT RISK SCORES

| No. | Case number | Heart risk score |
|---|---|---|
| 1 | 86 | 168.12 |
| 3 | 29 | 150.5 |
| 5 | 97 | 145.73 |

To get a more comprehensive picture of the influence of each predictor on the top three AR cases, the value of each has been presented in table 7 below. We note that the average variation in the risk scores using 5-fold cross validation is 0.0015, which is clinically negligible.

The values of predictors for Case No. 86, 29, and 97 are as follows (refer to Table 7):

**TABLE 7**
DISTRIBUTION OF HIGH-RISK PREDICTORS IN THE TOP THREE AR CASES

| No | Predictors | Top three at-risk cases | | |
|---|---|---|---|---|
| | | 86 | 29 | 97 |
| 1 | Age | 67 | 65 | 62 |
| 2 | Sex | 0 | 0 | 0 |
| 3 | cp | 2 | 2 | 0 |
| 4 | trtbps | 115 | 140 | 140 |
| 5 | chol | 564 | 417 | 394 |
| 6 | fbs | 0 | 1 | 0 |
| 7 | restecg | 0 | 0 | 0 |
| 8 | thalachh | 160 | 157 | 157 |
| 9 | exng | 0 | 0 | 0 |
| 10 | oldpeak | 1.6 | 0.8 | 1.2 |
| 11 | caa | 0 | 1 | 0 |
| 12 | thall | 3 | 2 | 2 |

## 4. Discussion

This section examines the rationale behind the high-risk scores computed for the aforementioned cases, emphasizing the plausible influence of key predictors. Table 6 establishes that females aged 62–67 years (mean age 64.66 years) with hypercholesterolemia (mean level 458.33 mg%), recorded a high maximum heart rate ('thalachh' with a mean of 158 beats per minute) and elevated values in the thallium stress test ('thall' with a mean of 2.33). This set is at the highest risk of heart attack for this population. Notably, these individuals exhibit no significant changes in resting ECG, no exercise-induced angina, and no apparent coronary vessel involvement or blockades. Furthermore, the 'old peak' parameter, indicating ST depression induced by exercise relative to rest, remains unaltered in these cases. *These findings are clinically intriguing and warrant further investigation.*

The objective of this study is to design, develop, and deploy a clinical decision support system and to statistically measure the performance of the algorithms. Silhouette score or Silhouette coefficient, measuring the inter-cluster distance and compactness of datapoints grouped inside any cluster for each algorithm, is evaluated using equation 4 (refer to Table 3). Here, squared Euclidean is applied as the measure of distance. The Silhouette coefficient score varies between +1 and -1 where -1 is considered 'misclassifications' and +1 refers to best classification, respectively. The score '0' or close to '0' indicates a chance of overlapping clusters. In the table, the Silhouette score DBSCAN method cannot be calculated as it failed to cluster the dataset into AR and NAR groups. GMM, K-means, BIRCH, and Affinity propagation algorithms show comparable scores, where the coefficient score of GMM is the best. Clinically, GMM can also cluster the datapoints more accurately when validated based on the class labels of the working dataset. However, the Silhouette score statistically indicates some overlapping of the classes and some degrees of dispersed datapoints, which need to be revisited. We note that any Silhouette score over 0.25 can be accepted as reasonable [25]. For this study, it is 0.2623, which is greater than the acceptable threshold. Hence, the chance of some degrees of overlapping and non-coherence among the datapoints within the clusters are accepted. Although the accuracy measure has been given more importance than the Silhouette coefficient score as the goal of any CDSS for clinical consistency, technically the algorithm must also be sound. In this case, GMM shows both clinical and technical soundness, which is the ideal condition.

The higher prevalence of heart attacks in *postmenopausal women* compared to men of similar age is well-documented [26] and as mentioned above, women with mean age of 66.64 years usually fall under postmenopausal group. This corroborates the cardioprotective role of estrogen in premenopausal women is attributed to its ability to reduce oxidative stress by facilitating the production of antioxidative enzymes and clearing reactive oxygen species [27], which is affected in the postmenopausal women. *Elevated blood cholesterol* levels increase heart attack risk by promoting atherosclerotic plaque formation and coronary vessel blockage [28]. However, postmenopausal women often experience heart attacks without significant vessel involvement [29], which could be linked to the lingering protective effects of estrogen on vascular endothelial cells. These effects may involve the clearance of deposit granules through autophagy, mediated by estrogen receptor-alpha activation [30]. A plausible mechanism for these heart attacks is *coronary vasospasm* rather than atherosclerotic blockades, as the loss of estrogen's vasodilatory effect may trigger sudden coronary spasms in vulnerable cases [29, 31].

*Anxiety* and *stress*, which are known to increase after menopause, contribute to elevated resting heart rates, frequent palpitations, and heightened heart attack risk in middle-aged and elderly women [32]. Between 15% and 50% of postmenopausal women experience anxiety disorders and depression [33], potentially due to the abrupt depletion of estrogen receptors in the brain [34]. This decline occurs more dramatically in cases of artificial menopause (e.g., surgical, chemical, or radiation-induced) compared to natural menopause [35]. The resulting mismatch between myocardial oxygen demand and supply may lead to early-stage heart muscle fatigue. However, many at-risk individuals initially present with normal stress test results and no detectable ECG abnormalities [31]. However, the decline of estrogen receptors and thus lower estrogen response in the body and mind (the physiological intrinsic factor) is still not conclusive as not every postmenopausal woman experience similar degrees of anxiety and stress during menopause. We also feel that other intrinsic factors, such as genetic predispositions, family history are some important components behind this variation. The difference could also related to biological variations in the rate of estrogen decline and/or influenced by various extrinsic modulators (i.e., environmental factors), such as unemployment or fear of job loss, severe illness, hospitalization and its respective financial and social consequences, loss of life and so forth. Any or all of these may have played critical roles during COVID-19 pandemic at the backdrop of emotional stress and anxiety. Thus, inclusion of the extrinsic stress factors as the predictors could be appropriate to model heart attack risk in postmenopausal women. It is also worth noting that during the pandemic period, emergency economic policies were rolled out to curb the financial stress and anxiety in society [36, 37]. Inclusion of these factors may provide a more holistic and balanced heart attack risk model in the context of COVID-19.

Applications of artificial Intelligence (AI) and ML are being used to cater to quality healthcare to a larger population in line with 'health for all'. These technologies are used for predicting diseases and conditions with higher precision typically complementing, or at times, superseding human intervention. This is an evolving field of research [38, 39] with key inroads already made in the fields of pathology and radiology [40], as also in oncology [41] research. Unsupervised machine learning techniques, particularly clustering, offer valuable tools for identifying patterns in unlabeled patient populations, as in this study. This approach is especially beneficial in subjective clinical decision-making scenarios where patient conditions cannot be readily classified during early stages, such as in psychiatric and psychological disorders. Clustering methods have been widely applied to group depression cases based on patterns [42], classify psychotic disorders [43], and evaluate clustering performance across domains [44, 45]. Among these methods, Gaussian Mixture Model (GMM) clustering has distinct advantages over approaches such as k-means clustering. As a Gaussian probabilistic method, GMM provides a reliable density estimation, accounts for uncertainty within groupings, yields opportunity for a datapoint's measurable inclination to be grouped into other clusters as well, and as a result, allows for overlapping clusters. By assigning a probability distribution to each cluster, GMM offers greater flexibility and granularity in cases where cluster boundaries are ambiguous [46], which is a natural occurrence with the clinical and biological datasets. This clustering method is meaningfully versatile and has been effectively utilized in identifying key protein candidates for drug design [47], stratifying clinical data into high-risk and low-risk groups of heart attack [13], scoring recurrence risks in patients of differentiated thyroid cancers [48], and population segmentation based on genomic datasets [49]. The unequivocal conclusion that can be drawn from this study using a hybrid machine learning method is that *menopause* in women is a critical physiological threshold aggravating the risk of heart attack. While this was not unknown previously, the precise dependence of heart attacks on menopause is an altogether new finding from this study, one that can be an efficient guideline in clinical decision-making.

*Major Outcomes*:

a) We introduce unsupervised learning techniques, specifically clustering methods, to effectively analyze patterns in "unlabeled" data on heart attack risk, enabling the identification of high-risk subpopulations. The aim is to design, develop, and deploy a CDSS on heart risk data.

b) The study identifies the optimal clustering technique by accurately grouping similar members within clusters while excluding dissimilar members, ensuring robust categorization. In this work GMM stands superior to its competitors.

c) Cross-correlation coefficients ('r' values) are computed for the features of the high-risk clusters to quantify and validate the strength of their positive relationships and these are statistically significant.

d) High correlation values ('r' values) are utilized as coefficients in the construction of a linear regression model (refer to Equation 2), where the predictors correspond to the identified variables.

e) Using the regression model, heart attack risk scores are computed on a case-by-case basis, providing individualized risk assessment.

f) The distributions of the predictors are explained, and the characteristics of high-scoring cases are thoroughly discussed to provide deeper insights into the risk factors.

g) The computed risk scores inform targeted care management strategies aimed at preventing heart attacks within vulnerable populations. The lateral outcome of this study is the extraction of importance-scores associated with all risk factors. For a practicing clinician, this could provide precise leads, e.g. how a 10% increase in cholesterol for males in the age range 50-55 could exacerbate heart attack risk compared to the age range 60-65 for females.

*Limitations of the study*:
   a) The sample size (N = 303) is small.
   b) Not all features, such as lifestyle-related predictors, e.g., tobacco smoking, sedentary lifestyle, alcoholism, and other substance abuse, etc. could be considered due to lack of appropriate data [50].
   c) Risk validation by the medical doctors must involve data from 'real-world' cardiac risk datasets, which opens the scope of future work.
   d) The study does not establish any direct link between heart health and COVID-19 pandemic and vaccination effects. Plausible associations are assumed based on the published documents.
   e) The model could be more robust if other extrinsic and intrinsic factors could be associated, and
   f) Although GMM can cluster the dataset into AR and NAR classes most accurately, yet its Silhouette coefficient value (0.2623) indicates some overlapping between the classes. This needs to be revisited.

5. **Conclusions**

This study aims to identify high-risk heart attack cases from the publicly available Kaggle heart attack risk dataset using five clustering techniques and to evaluate their comparative performance. Amongst these, GMM demonstrated superior clustering ~~efficacy,~~ accuracy identifying 84.24% of high-risk cases, with an overall accuracy rate of 86.68%. Although, technically, there is some overlapping between the clusters, accuracy in classifying has been given priority from the clinical perspective. However, such clinic-technical discrepancy may warrant a rethink in future from the design, development, and deployment of any CDSS using clustering algorithms. The analysis further assesses heart attack risk by calculating Pearson's correlation coefficients ('r' values) among the predictors for each cluster. The correlation values are statistically significant. Notably, the maximum heart rate achieved post-exercise ('thalachh') and the slope of the ECG ('slp') exhibit the strongest positive correlation (r = 0.38). These high 'r' values are incorporated into a multiple linear regression model, where they serve as the predictors whose coefficients are computed to fit the information into Equation (2).

The findings reveal that middle-aged and elderly women (falling supposedly into postmenopausal group, at least physiologically speaking) are at the highest risk of heart attack, potentially due to the loss of estrogen, which has a 'cardioprotective' as well as 'vasoprotective' effect. Based on this result, the study highlights the importance of increased vigilance in identifying menopausal and perimenopausal women as high-risk groups for heart attacks, paralleling the typical focus and emphasis which is typically put on men. Estrogen could be the key player behind the elevated risk of heart attacks in postmenopausal females as stated above. Also, extrinsic factors, e.g., adverse socio-economic situation during the pandemic, relevant financial policies, and emergency measures encumbering all these elements may contribute to alleviation of socioeconomic stress that need to be considered for a more robust model of heart attack.

**Author Contributions:** Conceptualization, S.C.; methodology, S.C. and A.K.C.; software, S.C.; validation, S.C. and A.K.C.; formal analysis, S.C..; investigation, S.C..; resources, S.C..; data curation, A.K.C.; writing - original draft preparation, S.C..; writing - review and editing, A.K.C.; visualization, S.C.; supervision, S.C. and A.K.C.; project administration, S.C. All authors have read and agreed to the published version of the manuscript.

**Funding:** This research received no external funding.

**Data Availability Statement:** The study used freely available (open access) Kaggle data that can be accessed from [14]. Data for the plots and tables can be made available by the corresponding author on request.

**Acknowledgments:** S.C. acknowledges the open access data archive of Kaggle.

**Conflicts of Interest:** Authors declare no conflicts of interest.

**Abbreviations**

The following abbreviations are used in this manuscript:

| | |
|---|---|
| AI | Artificial Intelligence |
| AR | At Risk |
| CDSS | Clinical Decision Support System |
| COVID | Corona Virus Disease |
| CVD | Cardiovascular Disease |
| GMM | Gaussian Mixture Model |
| HRV | Heart Rate Variability |
| ML | Machine learning |
| NAR | Not At Risk |